\documentclass[referee]{aa} 
\usepackage{graphicx}

\newcommand{\lsim}{{\, \lower2truept\hbox{
${< \atop\hbox{\raise4truept\hbox{$\sim$}}}$}\,}}
\newcommand{\gsim}{{\, \lower2truept\hbox{
${> \atop\hbox{\raise4truept\hbox{$\sim$}}}$}\,}}

\begin{document}
\baselineskip=12pt
   \title{ The Evolution of Nova V382 Vel 1999 }


   \author{M. Della Valle
          \inst{1}
	  L. Pasquini
	  \inst{2}
	  D.Daou	
          \inst{3}
          \and
          R. E. Williams\inst{4}
          }


\institute{Osservatorio Astrofisico di Arcetri, Largo E. Fermi 5, Firenze, 
Italy\\ 
\and
European Southern Observatory; KarlSwarschildstrasse 2, Garching bei 
M\"unchen, M\"unchen, Germany\\ 
\and
California Institute of Technology MS: 220-6 1200 East California Blvd.,
Pasadena, CA 91125\\
\and 
Space Telescope Science Institute, 3700 San Martin Drive,
Baltimore, MD, USA\\ 
}


\date{Received --; accepted --}

\abstract{ We report results of spectroscopic observations of V382
Vel (Nova Vel 1999) carried out at La Silla between 5 and 498 days
after maximum light (23 May 1999, V(max) $\sim 2.3\pm 0.1$).  The
analysis of the photometric and spectroscopic evolution shows this
object to be a {\sl fast nova} belonging to the Fe II {\sl broad}
spectroscopic class. A distance of 1.7 kpc ($\pm 20\%$) is derived
from the maximum magnitude vs. rate of decline relationship after
correcting for the small reddening toward the nova, E(B--V)$\lsim
0.10$. From the measured H$\alpha$ flux and the associated rate of
expansion we derive an approximate mass for the ejected shell,
M$_{env}\lsim 10^{-5}$ M$_\odot$. We have also observed during the
early decline a broad, short-lived ($\lsim$ 10 days) feature at
6705-6715 \AA~ for which several identifications are possible, one of
which is the lithium doublet at 6708 \AA~ and which could place an
empirical limit on the lithium production that might occur during the
outburst of a {\sl fast nova}.
The high luminosity at maximum, M$_v=-8.9$, and the relatively small
height above the galactic plane ($z\lsim 160$pc) suggest that V382
Vel originated from a massive white dwarf, likely in the mass range
1.1--1.2 M$_\odot$.

\keywords{Cataclysmic Variables --
          Nova stars 
               }
}   
\authorrunning{ }

   \maketitle

\section{Introduction}

Nova Velorum 1999 (=V382) was independently discovered by P. Williams
(1999) and A. Gilmore (1999), on May 22, as a star brighter than
V$\sim 3$, located at R.A. = 10h44m48s.39, Decl. = --52o25'30".7
(equinox 2000.0, Garradd 1999).  Early spectroscopy obtained by Lee et
al. (1999) confirmed the object to be a classical nova undergoing
outburst.  V382 Vel achieved optical maximum roughly 1 day
later. This brightness makes V382 Vel the brightest
galactic nova since V1500 Cyg 1975 and one of the brightest novae of
the century (see Table 1), thus offering us a unique opportunity to
study its photometric and spectroscopic evolution in great detail.
Its spectroscopic evolution was followed at La
Silla Observatory as an ESO target of opportunity with the 1.5m
telescope equipped with FEROS (Fiber-fed Extended Range Optical
Spectrograph), a high-resolution echelle
spectrograph characterized by high efficiency and large
wavelength range, covering the entire optical spectral range
(3800-9000 \AA~) in a single format at a resolution of R=48000.

\begin{table}
\begin{center}
\caption{Brightest Novae since 1900} 
\begin{tabular}{cccc}
\hline

\# & Nova & Year & V(max)\\
\hline

 1. & v603 Aql & 1918 &             --1.1 \\
 2. & GK Per   & 1901 &              0.2\\
 3. & CP Pup   & 1942 &              0.5\\
 4. & RR Pic   & 1925 &              1.0\\
 5. & DQ Her   & 1934 &              1.3\\
 6. & v476 Cyg & 1920 &              1.6\\
 7. & v1500 Cyg& 1975 &              1.8\\
 8. & CP Lac   & 1936 &              2.1\\
 9. & {\underline {V382 Vel}} & 1999 &              2.3\\
10. & v533 Her & 1963 &              2.5\\
11. & Q Cyg    & 1876 &              3.0\\ 
    &v446 Her  & 1960 &              3.0\\

\hline  

\end{tabular}
\end{center}
\end{table}

\section{Lightcurve} 

The lightcurve of the nova and its color evolution are shown in Figs. 1a
and 1b using the photometric measurements of Gilmore and
Kilmartin (1999) with the 0.6m telescope at Mount John University
Observatory.  An extrapolation of the photometric measurements near
maximum light indicates that V382 Vel reached V=$2.3\pm 0.1$ on 23 May
1999.  The nova decreased in brightness by two magnitudes from maximum
in t$2=4.5$ days (t$3=9$d). Applying these data to the maximum
magnitude {\sl vs.} rate of decline relationship derived by Della
Valle and Livio (1995) for novae in M31 and the LMC indicate that V382
Vel was intrinsically luminous, achieving an absolute magnitude at
maximum of M$_V=-8.9\pm0.17$.  The observed (B--V) obtained by
averaging 4 measurements at maximum light yields $\langle$ (B--V)
$\rangle$ =0.28$\pm 0.03$. This is very similar to the value
(B--V)=0.23$\pm 0.06$ obtained by van den Bergh and Younger (1987) for
a sample of unreddened galactic novae, suggesting that V382 Vel
suffered little galactic absorption ($A_V\lsim 0.1-0.15$ mag).
Similar conclusions were drawn by Shore et al. (1999) after noting the
lack of the 217.5-nm interstellar feature in STIS spectra of the
nova. The evolution of the colors is very typical, becoming
bluer with time as the ejecta become optically thin and
the photosphere recedes. The smoothness of the lightcurve
during decline suggests little dust formation (if any) in the ejecta.
\smallskip

The mean magnitude of the nova at quiescence has been determined
by Platais et al. (1999) by examing photographic plates from the
Yale/San Juan Southern Proper Motion program. They derived V=$16.56\pm
0.13$ and an average (B--V)=0.14, which is consistent with the
unreddened colors of novae at minimum (cf. Bruch 1984). The
outburst amplitude of $\Delta M\sim 14.3$
magnitudes is larger than the average outburst for galactic
novae ($\sim 12$ mag), but is not unusual for {\sl fast novae} (for
example, CP Pup 1942 and V1500 Cyg 1975 exhibited amplitudes of
$\gsim 16$ mag and $\gsim 19$ mag, respectively). This amplitude
and the derived absolute magnitude at maximum lead to
M$_V\sim 5.3$ at minimum, which falls in the faint tail of the
distribution for classical novae (see Della Valle and Duerbeck 1993,
their Fig. 3).
\smallskip

Platais et al. (1999) also point out the discovery on a photographic
plate obtained on 28 April 1970 of a flaring of 0.5 magnitudes. Such
flaring is not uncommon in the years before a major outburst in
classical novae, e.g. the $\sim 0.5$ mag flaring detected in AR Cir
1906 five years before its major outburst (Fig. 2).

\section{Spectroscopic Observations}

\subsection{The early decline}

Spectroscopic observations of the nova began at ESO five days past
maximum as part of a target of opportunity campaign (Della Valle,
Pasquini and Williams 1999).  We have for the first time exploited the
potential of the ESO 1.5m telescope equipped with FEROS (see Kaufer et
al.  1999) on a bright galactic nova, and have collected a series of
13 spectra in the optical range 3800-9000 \AA~ with a resolution of
48000. Five representative spectra are shown in Fig. 3a and 3b. The
first spectrum, obtained on 28 May, exhibits a bluish continuum
dominated by Balmer emission, Fe II and other low ionization heavy
element lines, in addition to two absorption systems, the {\sl
'Principal'} and {\sl 'Diffuse Enhanced'} absorption normally observed
in early decline (Payne-Gaposchkin 1957). The expansion velocities for
these systems measured from the minima of the Balmer line P-Cyg
profiles are close to 2300 and 3700 km/s respectively, and the HWZI
measured for the H$\alpha$ emission feature is 3600 km/s.  H$\alpha$
and H$\beta$ show rather asymmetric `saddle-shaped' profiles (see
Fig. 4a and 4b), with the blue component clearly more prominent than
the red. This difference disappears with time, and between 3 and 6
June the blue component diminishes, as the profiles evolve to a
flat-topped shape.  At the same time the steady decrease in the
expansion velocity of the emitting gas (Fig. 5), initially decreasing
as a power law in time [$\sim \Delta t^{n}$ with $n$ varying from
--0.32 (H$\alpha$) to --0.15 ($H\beta$) and --0.10 (H$\delta$),
$\Delta t$=time since maximum light], changes character as it reaches
$\Delta t \sim 10$ days slowly decreasing during the following $\sim
400$ days by $\sim 10-15\%$ only. During this interval (the initial 15
days past maximum) the nova decreases in brightness by more than 3
magnitudes from maximum.  The identifications and relative intensities
of all emission lines observed in the spectra between 28 May 1999 and
2 Oct 2000 are given in Table 3. The line identifications are provided
in col. 1, whereas the intensity of each line has been determined,
independently by two of us, by fitting its profile (or deconvolved
profile in the case of blends) with multiple gaussian fits in MIDAS
(Banse et al. 1988) (using the {\sl Alice package}) and IRAF
environments.  Intensities are given relative to H$\beta$
(cols. 2--14).  The FEROS spectra were not calibrated in absolute flux
on the nights that observations were obtained, so a relative FEROS
response curve was obtained from the spectra of spectrophotometric
standards observed during commissioning (Kaufer et al. 1999).  The
high stability of the instrument and the large diameter fibre aperture
(2.5 arcseconds) result in a stable instrument response function that
can be used even though not obtained on the same nights as the
observations.  Absolute fluxes were determined by matching the
observed energy distributions with the corresponding $B$ and $V$
photometric measurements (cf. Table 2).  The most striking change in
the spectrum occurs one month after maximum light when the early low
ionization permitted transitions give way to emerging higher
ionization and forbidden lines.  This is accompanied by a change to
flat-topped line profiles, probably associated with the termination of
the post- outburst wind phase and the complete ejection of the
envelope.

\subsection{\sl The nebular stage}

The nova entered the nebular spectral stage by the end of June, roughly
40 days after maximum light. At that time the spectra were
characterized by the increasing strengths of [O III] $\lambda \lambda
4959, 5007 \AA~$ and He I lines, and the gradual disappearance of the Fe II
lines and the absorption systems. The spectra obtained on Nov 22,
1999 and Oct 2, 2000 show [O III] 
brighter than H$\alpha$+[N II] and exhibiting a flat-top, castellated
structure with a HWZI of $\sim 2000$ km/s, with the He
I lines fading and soon disappearing. During this stage we note
also the emergence of strong [Ne III]
$\lambda \lambda 3869, 3968\AA~$, and [Fe VII] $\lambda \lambda 6087 \AA~$,
which are typical of the nebular
stages of novae belonging to the He/N and Fe II `broad' spectroscopic
class (see Williams 1992).

\begin{table}
\begin{center}
\caption{ H$\beta$ Fluxes: erg cm$^{-2} s^{-1}$}
\begin{tabular}{cccc}
\hline

Date & $\Delta t$ & F$_{H\beta}$ & \% error\\
\hline

          28 May 99  &       5   & 2.54e-11 &  0.39\\
          29 May 99  &       6   & 2.20e-11 &  0.30\\
          30 May 99  &       7   & 1.54e-11 &  0.25\\
           1 Jun 99  &       9   & 1.13e-11 &  0.21\\
           2 Jun 99  &      10   & 8.96e-12 &  0.17\\
           3 Jun 99  &      11   & 1.02e-11 &  0.28\\
           5 Jun 99  &      13   & 5.34e-12 &  0.42\\
           6 Jun 99  &      14   & 3.50e-12 &  0.29\\
          25 Jun 99  &      33   & 2.89e-12 &  0.06\\
          14 Jul 99  &      53   & 1.82e-12 &  0.07\\
          19 Jul 99  &      58   & 1.63e-12 &  0.37\\
          31 Jul 99  &      69   & 1.07e-12 &  0.22\\
          22 Nov 99  &     183   & 3.89e-13 &  0.45\\
	   2 Oct 00  &     498   & 5.62e-14 &  0.30\\
\hline  

\end{tabular}
\end{center}
\end{table}

\section{Interstellar extinction}

As previously mentioned the nova showed little evidence for
extinction near maximum light. An independent estimate of the
reddening at subsequent times can be made from a comparison of
predicted and observed emission lines ratios within the same ion
which are not sensitive
to radiative transfer effects (Robbins 1968, Ferland 1977).
Unblended H and He recombination
lines may be used for this purpose if they are optically
thin and not affected by collisions. Because H$\alpha$ is seriously
blended with the [NII] lines the HeI lines offer the best possibility
for this purpose, and data presented in Table 3 provide ratios at 6
epochs with a median value of $\lambda5876/\lambda4471=3.06$, which
can be compared with the recombination value of 2.9. Using
the ratio of total to selective absorption R=3.2 (Seaton 1979),
one derives $E_{B-V}=0.05$ (neglecting collisional
effects).  A separate value of reddening can be derived from the
$\lambda6678/\lambda4471$ ratio. The observed value (median from 5
epochs) of 0.77 is close to the theoretical value of 0.78.

All nova spectra show clear insterstellar NaI D and Ca II H and K
absorption.  Na D1 and D2 are saturated, and the minimum intensity of
the lines is zero intensity, however the Ca II H line does show some
residual intensity. For both doublets only one absorption component is seen
at our resolution. The equivalent widths of the lines as measured from our
FEROS spectra are 360 and 395 m$\AA~$ for the D1 and D2 lines, and
203 and 382 m$\AA~$ for H and K of CaII. The IS component has a
velocity of 4.4 km/s and the width of the lines (given as the FWHM
of the gaussian fitting of the different lines), is 0.23 and 0.25
$\AA~$ for Ca II H and K (17.4 and 19.1 km/s), and 0.322 and 0.345
angstrom for the D1 and D2 components (16.4 and 17.6 km/s). By
applying crude estimates for the reddening vs. Na D equivalent widths,
e.g. E(B--V)=$0.13 \times$ EW(D lines)
(Benetti, private communication) we find E(B--V)$\sim 0.09$.
Although this scaling relation is approximate, a
comparison with the Na EW observed in detailed IS studies with
resolved single components [e.g. the SN 1987a field Molaro et
al. (1993)] indicates a reddening slightly higher than that
estimated from emission line ratios, viz., $A_V\lsim
0.3$ mag.  In general, however, all methods confirm a rather low
reddening and therefore a moderate distance to the nova.

\section {\bf {The [O I] and [O III] lines}}

Measurements of the intensities of emission lines in nova spectra
have shown an anomalous ratio of the [O I] $\lambda \lambda
6300, 6363 \AA~$ nebular lines (see Williams 1994). The relative
intensities of these lines is usually not 3:1 (an exception
was FH Ser 1970, see Rosino et al. (1986)), as expected from their atomic
transition probabilities. Nova Vel 1999 is no exception as the data
shown in Table 3 averaged over 11 epochs lead to
a $\lambda$6300/6363 ratio of $1.8\pm 0.3$.  Williams (1994) has
discussed this situation in novae, and has suggested that optically
thick [O I] lines must be formed in very dense (n(H)$\gsim
10^{14} {\rm cm}^{-3}$), small blobs of neutral material embedded within 
the ionized ejected shells. 

During early decline when the density of the ejecta is in the
high-density limit of the [O I] lines the intensity ratio $F_{\lambda
6300}/F_{\lambda 5577}$ can be used to determine the temperature
(e.g. Osterbrock 1989) from the relation:
$$F_{\lambda 6300}/F_{\lambda 5577}=0.023 \times (1-e^{-\tau})/\tau
\times {\rm exp}(25800/T_e)$$ 
The [O I] line at $\lambda 5577$ is clearly
detected in the spectra of on 5 and 6 June with 
$F_{\lambda 6300}/F_{\lambda 5577}
=2.19$. This is within the range of measured intensity
ratios reported by Williams (1994), although close to
the lower limit of the sample, and implies $T_e(K)=4800$.

The evolution of the nebular and auroral [O III] emission lines,
$F_{\lambda(4959+5007)}/F_{\lambda4363}$, ($\sim 1$, on 31 July 1999;
$\sim 3.1$, on 22 November 1999; and $\sim 36.7$ on 2 October 2000),
suggests (cf. Filippenko and Halpern 1984) for the typical range of
temperatures $3.4\lsim Log T_e \lsim 3.7$ (see Williams 1994, his Tab. 3),
values of the density initially (31 July) close to the high density limit
$n_e\sim 10^9 {\rm cm}^{-3}$, and eventually decreasing to $n_e\sim 10^4 
{\rm cm}^{-3}$.
\newpage

\begin{figure}
\centering
\includegraphics{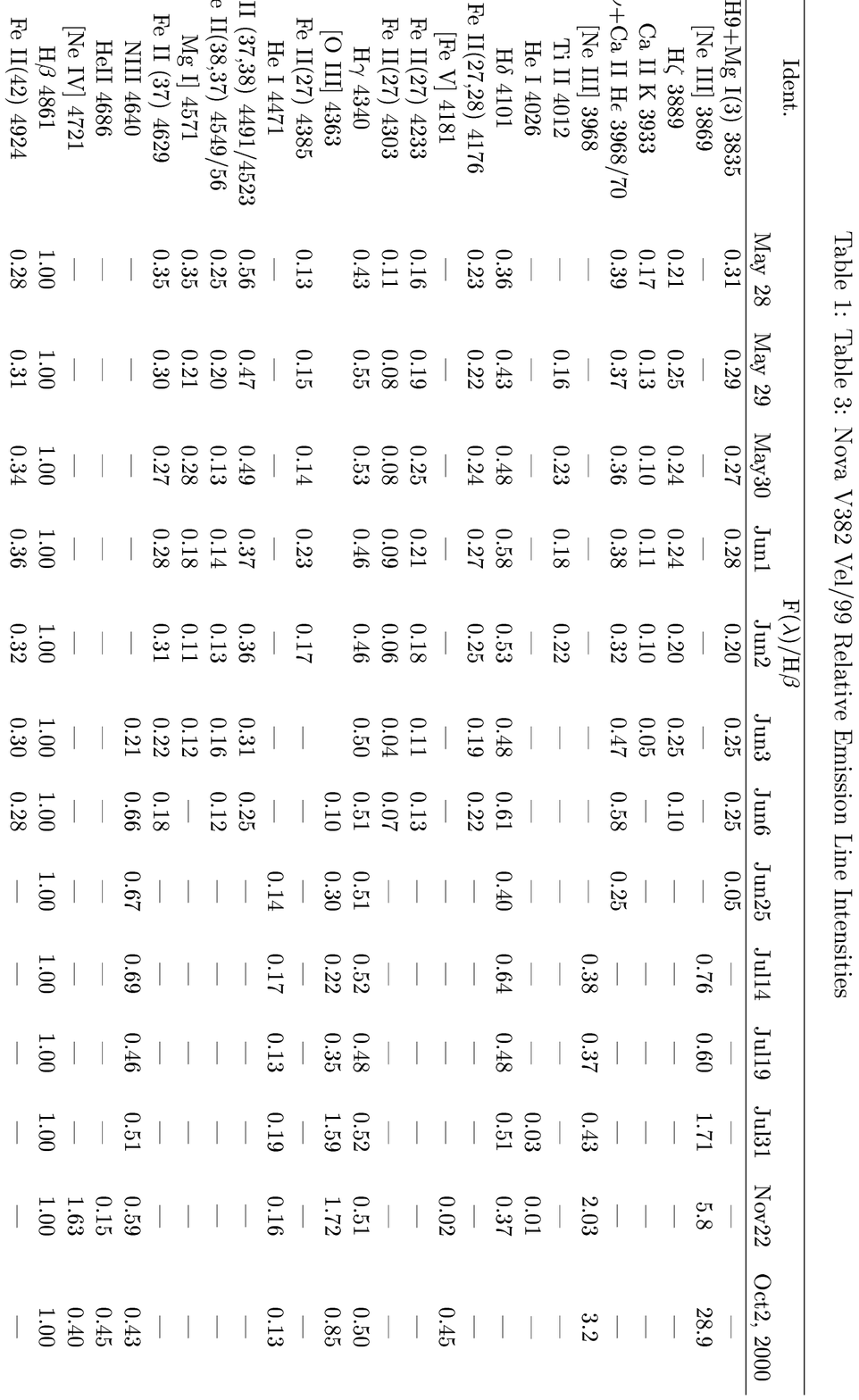}
\end{figure}

\newpage

\begin{figure}
\centering
\includegraphics{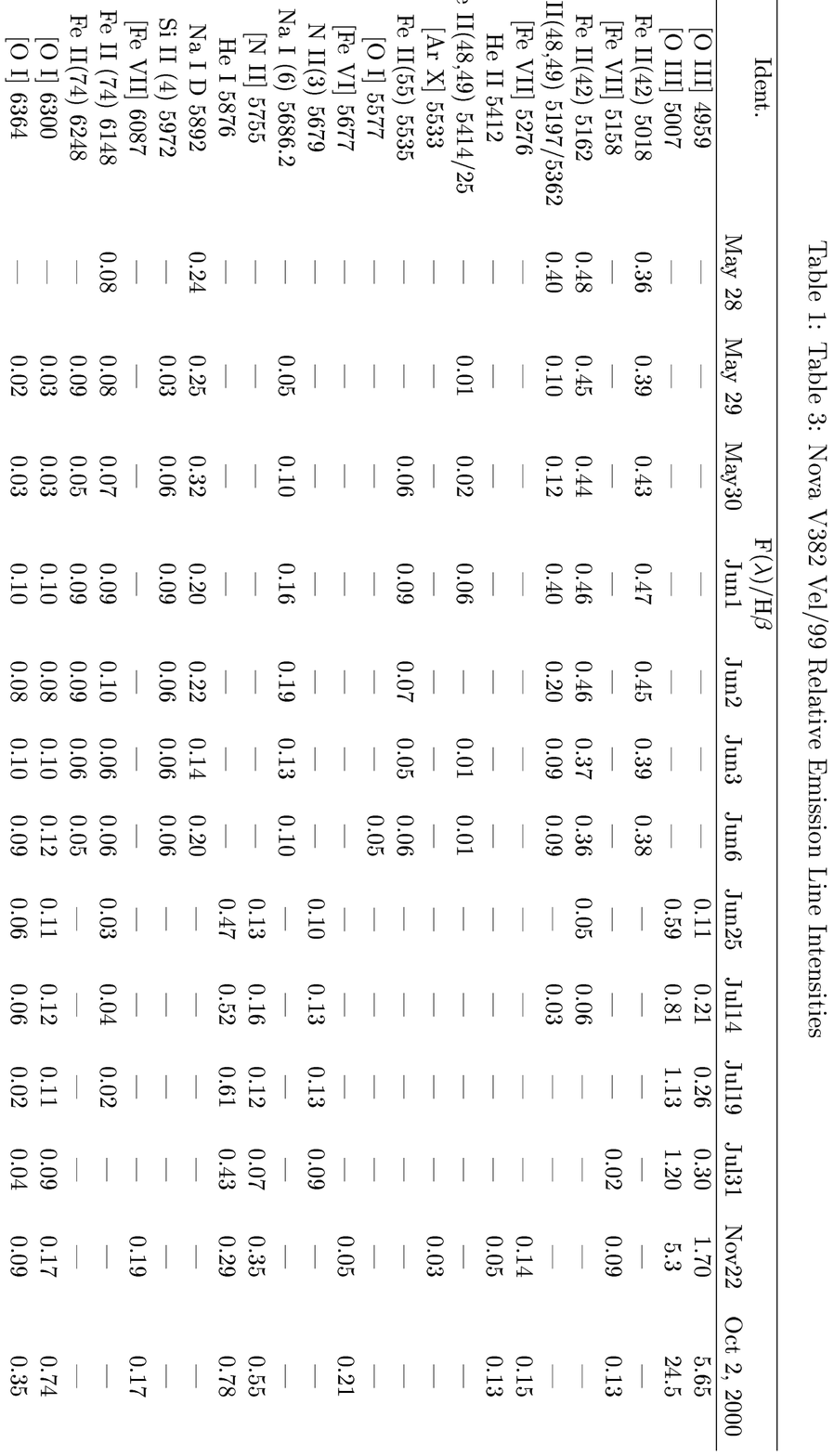}
\end{figure}

\newpage

\begin{figure}
\centering
\includegraphics{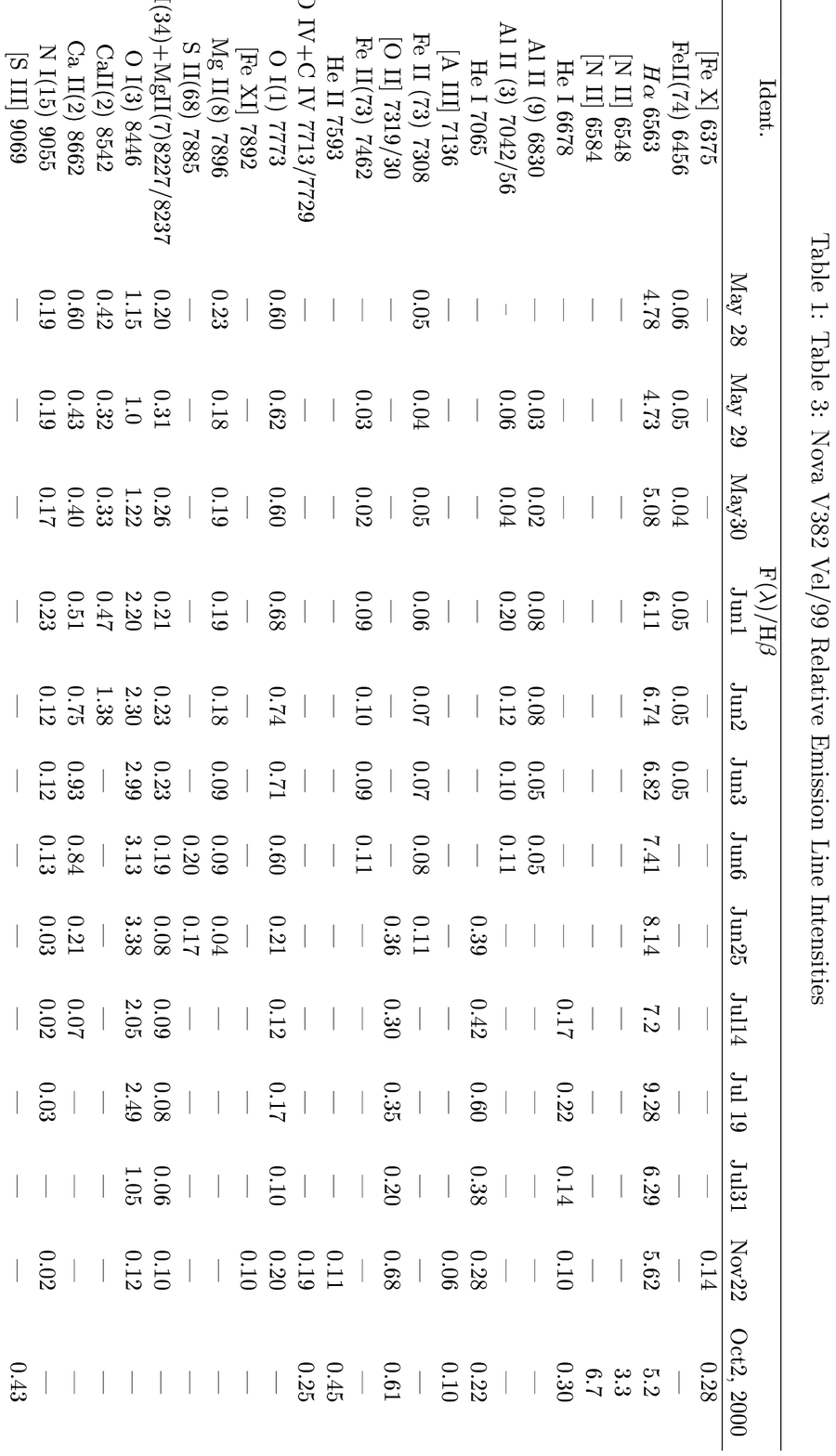}
\end{figure}

\section{The mass of the ejected envelope}

The most reliable distances to classical novae are parallaxes derived
from angular size measurements of the expanding ejecta.  Currently
this method can be applied to about twenty galactic novae (out of more
than 200 objects discovered to date), therefore most nova distance
measurements involve the use of the maximum magnitude vs. rate of
recline relationship. First noted by Zwicky (1936), it is
characterized by an intrinsic scatter of 0.16 mag ($1\sigma$), (Della
Valle and Livio 1995) corresponding to a distance uncertainty of $\sim
20\%$ ($3\sigma$). The photometric data presented in the introduction
yield a reddening-corrected distance modulus (m--M)=11.04, which
corresponds to a distance of $1660\pm 110$(1$\sigma$) pc.  The emitted
luminosity of H$_\alpha$ follows from the distance, from which an
estimate of the mass of neutral hydrogen content in the ejecta can be
made. The H$\alpha$ emission from the envelope is $E\alpha=g_\alpha
\times n_e^2 \times \epsilon \times Volume$ erg$^{-1}$ where $g_\alpha
\gsim 10^{-24}$ erg cm$^3$ s$^{-1}$ is the emission coefficient per
cm$^3$, for T$\lsim 5000$ (and $n_e=1$) and $\epsilon$ the volume
filling factor.  Assuming the volume of the expanding shell to be
$V\sim 4\pi \times R^2 \times \delta$, where R$\lsim v_{exp}\times
\Delta t$ ($\Delta t$=498 days from the maximum light and
$v_{exp}$=1600km/s) and $\delta =fR$ (with $f$ smaller than 1) is the
thickness of the shell. From the data of Tab 2 and 3 (2 Oct 2000), 
we obtain:
$$n_e^2= {7\times 10^6 \over \delta \epsilon}$$ If we assume the thickness
of the envelope increases from thermal motions inside the gas (Mustel
\& Boyarchuk (1970)), then for T$\sim 5000^\circ$ K, $f\sim
v_{ther}/v_{exp}\lsim 0.2$ and then $n_e\lsim 10^4 {\rm cm}^{-3}$,
which yields $6.5\times 10^{-6}$ M$_\odot$. This is an order of
magnitude smaller than ejecta mass determinations for {\sl slow novae},
but not unusual for {\sl fast} novae. Indeed being $\langle \Delta
m_{eject} \rangle \propto R^4_{WD}/M_{WD}$ and M$_B{\rm
(max)}\propto logM_{WD}$ (Livio 1992), the mass of the shell should
correlate with the rate of decline (in view of the existence of the
Maximum Magnitude {\sl vs.}  Rate of Decline relationship), as we 
actually observe, at $\sim 95\%$ confidence level (see Fig. 6)

$$ {\rm log~}\Delta m_{eject}=0.274(\pm 0.197)\times {\rm log} t_2 + 0.645(\pm 0.283)$$

However we note that the previous calculation refers to the H
ionized mass, therefore our estimate should be regarded as a lower
limit for the mass of the ejected envelope.

Observations of the [O I] line intensities formed when the
density in the ejecta exceeds critical density, after correction for
reddening and optical depth effects, yield the oxygen mass of the
neutral component into the ejected shell when the distance and the
electron temperature are known (e.g. Williams 1994). The previous data
give M$(O^0)\sim 9.3\times 10^{-8}$ M$_\odot$ which is only $\sim 1\%$
of the total mass of the shell. On the other hand , this should be
regarded as a lower limit for the total amount of oxygen present in
the ejecta (typically $\lsim 15\%$, see Gehrz et al. 1998), because
this procedure does not include the contribution of other O ions.

\section{The short-lived feature at 6705--6715 \AA}  

A close inspection of the early spectra reveals the presence at
$\sim 6700/6715$ \AA~ of a weak and broad emission line (see Fig. 7) which
increases in strength during the early decline and reaches
maximum intensity about 10 days past maximumm, disappearing within a
few days (undetectable on the Jun 25 spectrum.)  The measured FWZI
corresponds to an expansion velocity of $\sim 1100$ km/s, about one
third of the velocity measured for the Balmer emission lines. The
identification of this feature is not straightforward and its
narrowness suggests that it could be a component of the broader
feature that is immediately blueward.  The measured wavelength and
rapid evolution with time and its presence during early decline
require that it originates from a low ionization ion
(e.g. Si, N, Ni, Al). There is no unambiguous identification for this feature,
since no abundant ion has a strong transition at this wavelength, however
one possibility is the 6708 \AA~ doublet of {\sl Li I}.

Novae have been predicted to be sources of galactic lithium (e.g
Romano et al 1999 and reference therein, although see Travaglio et
al. 2001 for a different view), however attempts to detect lithium
(see Friedjung 1979) have been unsuccessful. 
According to Starrfield et al. (1978) and Boffin et al. (1993) (see
also Norgaard \& Arnould 1975) {\sl Li} should be produced by the
radioactive decay of $^7Be$, whose half-life time is 53 days, and therefore
it is not expected to be detectable during the earliest stages of a nova. This
constraint may be relaxed if one assumes that {\sl Li} begins to form
prior to the peak of TNRs and the observed
variation of the intensity of the line during the early nova evolution
is dominated by ionization effects in the ejecta.  

The observed intensity of the unidentified
feature with respect to the emission lines of hydrogen is 
$\gsim 10^{-6}$. Assuming an abundance of H in the ejected shell of
$\sim 10^{-5/-6}$ M$_\odot$ (see above) the unidentified
feature represents at most an injection of Li into the interstellar
medium of $\sim 10^{-11/-12}$ M$_\odot$, if it is formed by recombination.
If $10^4$ is the average number of outbursts
experienced by a nova during its lifetime (Bath\& Shaviv 1978), then
about $\sim 10^{-7/-8}$ M$_\odot$ could be assumed as an upper
limit to the possible production of {\sl Li} during the lifetime of a
fast nova.  This value is appoximately
the theoretical prediction of lithium produced by a nova during
its lifetime (D'Antona and Matteucci 1991, Romano et al. 1999).

\section{Conclusions}

According to the quantitative criteria defined by Della Valle and
Livio (1998) there are two classes of novae drawn from two populations
of progenitors:  
\medskip

a) {\sl Slow Novae} achieve an absolute magnitude of M$_B\lsim -7.5$
at maximum, their lightcurves exhibit a slow decline (often
characterized by secondary maxima) with t$_3 \gsim 20$ days (or t$_2
\gsim 12$ days), and they normally belong to the Fe IIn (n=narrow)
spectroscopic class defined by Williams (1992). They occur at
distances of up to $\gsim 1$ kpc above the galactic plane and are
probably related to the Pop II stellar population of the bulge/thick
disk, and are therefore associated with less massive white dwarfs,
M$_{WD}\lsim 0.9-1$ M$_\odot$;

b) {\sl Fast Novae} have lightcurves characterized by very bright peak
luminosities, M$_B\lsim -9$, followed by rapid, smooth declines
(t$_3 \lsim 20$ days, or t$_2 \lsim 12$ days), and they mostly belong to the
He/N or Fe IIb spectroscopic classes defined by Williams (1992). The
progenitors originate from a relatively old Pop I stellar population
of the thin disk/spiral arm, and therefore they are located
close to the galactic plane, typically $z\lsim 100-200$
pc. The associated white dwarfs are rather massive, M$_{WD}\gsim
0.9-1$ M$_\odot$.

Nova Vel 1999 was the brightest nova to occur in the Milky Way in the
last 25 years, and the rate of decline implies that this nova achieved
an absolute magnitude of M$_B\sim -9$, suggesting that it probably
originated from a massive white dwarf, with M$_{WD}\sim 1.15$
M$_\odot$, and likely originating from an O-Ne-Mg WD (see Shore et
al. 1999a,b).  We also note: {\sl a)} its location relatively close to
the galactic plane, z$\lsim 160$ pc; {\sl b)} the fast and smooth
decline (t2=4.5$^d$, t3=9$^d$), without the secondary maxima or sudden
variations in brightness (e.g. DQ Her 1934) that are indicative of
dust formation in the ejecta; {\sl c)} quick evolution to the nebular
stage ($\sim$ 40 days, i.e. from 1/3 to 1/10 of the time required by a
typical {\sl slow nova}; {\sl d)} the presence of strong Fe II lines,
characterized (to start from phase $\sim 10$ days, i.e.  at the time
when most envelope had been already ejected) by almost `flat-topped'
profiles and high expansion velocities ($\gsim 3500$ km/s at half of
FWZI), indicating that this nova was forming the emission lines in a
discrete shell, and {\sl e)} the presence during the nebular stage of
forbidden lines of high ionization, such as [Fe VII], [Fe X] and [Fe
XI].  All of this toghether points out that this nova belongs to the
{\sl broad} Fe II class (Fe IIb) of Williams (1992) scheme of
spectroscopic classification (see also Williams, Phillips and Hamuy
1994) and it qualify as one of the best examples of a prototypical
{\sl fast nova}.

The analysis of the spectroscopic data has shown the existence of a
short-lived ($\lsim 1$ week) emission feature in the
region 6700-6715 \AA~. There is no unambiguous identification for this
feature, and one possibility is the 6708 \AA~ lithium
doublet.  Novae are believed to be sources of Li for the galactic ISM,
and the intensity of this emission line implies 
$\sim 10^{-11/-12}$ M$_\odot$ as an upper limit to the
lithium produced by a {\sl fast nova} in a single outburst, a value
that is near the
theoretical predictions currently circulating in literature (see
Romano et al. 1999, Jose and Hernanz 1998 and references therein).

We have also estimated the mass of ionized hydrogen in the envelope
from the emitted flux in H$\alpha$, and under simple geometrical
assumptions we obtain $M_{env}\gsim 6.5\times 10^{-6}$ M$_\odot$ which
corresponds to a rather small mass envelope, although not atypical for
a {\sl fast nova}. There is also a considerable uncertainty in this
value due to the unknown filling factor. If the filling factor of the
envelope is close to 10$^{-2/-4}$ rather than $\sim 1$, as {\sl HST}
observations of the {\sl Recurrent Nova} T Pyx suggest (Shara et
al. 1997), then a large decrease in the mass of the nova ejecta will
result.  Evidence in favor of a moderately small value for the filling
factor is indicated by the high optical depth of [O I] lines for Nova
Vel 1999, which suggest that part of the ejecta are condensed blobs of
neutral material immersed in the ionized environment of the
shell. Our data (collected on 2 October 2000) may suggest 
a value for the filling factor as small as $\epsilon \approx 0.1$.
\smallskip

A second source of uncertainty is our estimate of the thickness of the
shell. In the past, the simple assumption $f\sim v_{ther}/v_{exp}$ has
led to underestimates of the thickness of the ejected shells for two
cases, DQ Her 1934 (Humason 1940) and FH Ser 1970 (Della Valle et
al. 1997) by a factor $\sim 3-4$. Correcting for this fact,
$M_{env}\sim 10^{-5/-6}$ M$_\odot$ may be a more appropriate estimate.
\medskip

The relatively small distance to V382 Vel (which partially explains the lack
of reddening toward the nova, in spite of the small value of its
galactic latitude) and the high expansion velocity of the ejecta, make
Nova Vel 1999 a suitable target for {\sl HST} observations,
within a short time. A systematic follow-up of the nova remnant, in
H$\alpha$ and [OIII] emission, coupled with simultaneous ground-based
high-resolution spectroscopic observations, would offer us the unique
opportunity to study the detailed evolution of the ejecta and
to provide a reliable estimate of the mass of envelope with greater
accuracy, and thus to properly evaluate the
contribution of classical novae to the galactic nucleosynthesis.

\section{Acknowledgments} The authors are deeply grateful to the La
Silla staff for having promptly activated the service-mode and
providing the spectroscopic follow-up of the nova. MDV thanks Space
Telescope Science Institute, where part of this work was done, for its
hospitality. The authors are also indebted with F, Matteucci,
S. Starrfield, P. Kilmartin and A. Gilmore for helpful discussions and with 
an anonymous referee for his remarks, which have improved the presentation
of this paper. 

\section{References}

Anupama, G. C., Duerbeck, H. W., Prabhu, T. P., Jain, S. K. 1992, A\&A, 263, 87\\
Banse et al. 1988, MIDAS Manual\\
Bath, G.T., Shaviv, G. 1978, MNRAS, 183, 515
Boffin H. M., Paulus, G., Arnould, M., Mowlavi, N. 1993, A\&A, 279, 173\\
Bruch, A. 1984, A\&AS, 56, 441\\
D'Antona, F., Matteucci, F. 1991, A\&A, 248, 62\\
Della Valle, M. 1988, from {\it An Atlas of Nova Lightcurves}, p. 198,
PhD Thesis, Padova University\\
Della Valle, M., Duerbeck, H. 1993, A\&A, 271, 175\\
Della Valle, M., Livio, M. 1995, ApJ, 452, 704\\
Della Valle, M., Gilmozzi, R., Bianchini, A., Esenoglu, H. 1997, A\&A,
325, 1151\\
Della Valle, M., Livio, M. 1998, ApJ, 506, 818\\
Della Valle, M., Pasquini, L., Williams, R.E. 1999, IAUC 7193\\
de Freitas Pacheco, J.A. 1977, MNRAS, 181, 421\\
de Freitas Pacheco, J.A., da Costa, R.D.D., Codina S.J. 1989,
ApJ, 347, 483 \\
Ferland, G.J. 1979, ApJ, 231, 781\\
Ferland, G. J., Williams, R. E., Lambert, D. L., Slovak, M., 
Gondhalekar, P. M., Truran, J. W., Shields, G. A. 1984, ApJ, 281, 194\\
Friedjung, M. 1977, in The Novae and Related Stars, 
edited by M. Friedjung, Astrophysics and Space Science Library, Vol. 65\\
Garrad, G.J. 1999, IAUC 7177\\
Gehrz, R.D. 1988, Ann. Rev. Astr. Astrophys., 26, 377\\
Gehrz, R.D., Truran, J. W., Williams, R.E. 1993\\
Gehrz, R.D.; Truran, J.W. Williams, R. E., Starrfield, S. 1998, PASP,110, 3\\ 
Gilmore, A.C. 1999, IAUC 7176\\
Gilmore, A.C., Kilmartin, P.M. 1999, IAUC 7238\\
Jos\'e J., Hernanz M., 1998, ApJ 494, 680\\
Kaufer et al. 1999, The Messeneger 95, 8\\
Hartwick, F.D.A., Hutchings, J.B. 1978, ApJ, 226, 203\\
Hassall, B. J. M., Snijders, M. A. J., Harris, A. W., Dennefeld, A.,
Cassatella, M., Friedjung, M., Bode, M., Whittet, D., Whitelock, P.,
Menzies, J., Lloyd-Evans, T., Bath, G. T. 1990, in Physics of
Classical Novae. Proceedings of Colloquium No.122 of the International
Astronomical Union, held in Madrid, Spain, June 1989. Editors,
A. Cassatella, R. Viotti; Publisher, Springer-Verlag, Berlin, Germany;
p. 202\\
Hjellming, R.M. 1990, Lect. Notes Phys., 369, 169\\
Humason, M.L. 1940, PASP, 52, 389\\
Lee et al. 1999, IAUC 7176\\
Livio, M. 1992, ApJ, 393, 516\\
Martin, P.G., 1989, in Classical Novae, eds. M.F. Bode, A. Evans, Wiley \&
Sons, Chichester\\
Norgaard, H., Arnould, M. 1975, A\&A, 40, 331\\
Payne-Gaposchkin, C. 1957, The Galactic Novae, North-Holland Publishing 
Company, Amsterdam\\
Platais,I., Girard, T.M., Kozhurina-Platais, V., Van Altena, W.F., Jain, R.K.,
Lopez, C.E. 2000, PASP, 112, 224\\
Pottasch, S. 1959, Ann. d'Astrophys., 22, 412\\
Prialnik, D., Kovetz, A. 1995, ApJ, 445, 789\\
Raikova, D. 1990, Lect. Notes Phys., 369, 163\\
Robinson, E.L., AJ, 80, 515\\
Romano, D., Matteucci, F., Molaro, P., Bonifacio, P. 1999, A\&A, 352, 117\\
Rosino, L., Benetti, S., Iijima, T., Rafanelli, P., Della Valle, M. 1991,
AJ, 101, 1807\\
Saizar, P., Starrfield, S., Ferland, G.J., Wagner, R. M., Truran, J.W., 
Kenyon, S.J., Sparks, W.M., Williams, R.E., Stryker, L. L. 1991, ApJ, 367, 310\\
Saizar, P., Starrfield, S., Ferland, G.J., Wagner, R. M., Truran, J.W., 
Kenyon, S.J., Sparks, W.M., Williams, R.E., Stryker, L. L. 1992, ApJ, 398, 651\\
Shara, M.M., Zurek, D.R., Williams, R.E., Prialnik, D.,Gilmozzi, R.,  
Moffat, A.F.J. 1997, AJ, 114, 258\\
Shore, S. et al. 1999a, IAUC 7192\\
Shore, S. et al. 1999b, IAUC 7261\\
Snijders, M. A. J., Batt, T. J., Roche, P. F., Seaton, M. J., Morton, D. C.,
Spoelstra, T. A. T., Blades, J. C. 1987, MNRAS, 228, 329\\
Starrfield, S., Truran, J.W., Sparks, W.M., Arnould, M. 1978, ApJ, 222, 600\\
Stickland, D. J., Penn, C. J., Seaton, M. J., Snijders, M. A. J., 
Storey, P. J. 1981, MNRAS, 197, 107\\
Taylor, A. R., Hjellming, R. M., Seaquist, E. R., Gehrz, R. D. 1988, Nature,
335, 235\\
Travaglio, C., Randich, S., Galli, D., Lattanzio, J., Elliot, L., M., 
Forestini, M., Ferrini, F. 2001, ApJ, 559, 909\\
van den Bergh, S., Younger, P.F. 1987, A\&AS, 70, 125\\
Williams, P. 1999, IAUC 7176\\
Williams, R.E., Gallagher, J.S. 1979, ApJ, 228, 482\\
Williams, R.E. 1992, AJ, 104, 725\\
Williams, R.E., Phillips, M.M., Hamuy, M. 1994, ApJS, 90, 297\\
Williams, R.E. 1994, ApJ, 426, 279\\
Woodward, C.E., Gehrz, R. D., Jones, T.J.; Lawrence, G. F. 1992,
ApJ, 384, L41\\  
\bigskip

\section{Caption}

Fig.1a and 1b. 

The lightcurve of v382 Vel (top) and its color evolution (bottom).
\medskip

Fig.2 

The pre-outburst and outburst lightcurve of AR Cir 1906. The $\sim 0.5$
mag flaring occurred about 5 years before the major outburst (Della
Valle 1988).
\medskip

Fig.3a,3b

The spectroscopic evolution of v382 Vel 1999. Spectra obtained (from top)
on May 28, Jun 6, Jun 25, Jul 31 and Nov 22, 1999.
\medskip

Fig.4a,4b

The evolution of the H$\alpha$ and H$\beta$ profiles with the time.
From top, 28, 29, 30 May, 3, 6, 25 Jun, 19, 31 Jul, 22 Nov 1999, and 2 Oct
2000. 
\medskip

Fig.5

The evolution of the FWHM of H$\alpha$, H$\beta$,  H$\delta$ with the time.
\medskip

Fig.6

The correlation between the rate of decline and the mass of the
ejected envelope. Data from Gehrz 1988, Snijders et al. 1987,
Hjellming 1990 (V1500 Cyg), Gehrz 1988, Stickland et al. 1981 (v1668
Cyg); Hjellming 1990, Hartwick and Hutchings 1978 (HR Del); Ferland et
al. 1984, Martin 1989 (DQ Her); Woodward et al. 1992 (v838 Her);
Pottasch 1959, Ferland 1979 (CP Lac); Gehrz 1988, Hassal et al. 1990
(GQ Mus); Williams and Gallagher 1979 (RR Pic); Hartwick and Hutchings
1978, Hjellming 1990, Della Valle et al. 1997 (FH Ser); Gehrz 1988 (LW
Ser); Raikova 1990 (LV Vul); Gehrz 1988 (NQ Vul); Gehrz 1988, Saizar
et al. 1991 (PW Vul); Taylor et al. 1988, Saizar et al. 1992 (QU Vul);
Gehrz et al. 1993 (QV Vul), de Freitas Pacheco 1977, Ferland 1979
(IV Cep), de Freitas Pacheco, da Costa \& Codina 1989 (v842 Cen);
Snijders et al. 1987 (v1370 Aql).
\medskip

Fig.7

The evolution of the broad feature at 6705-6715 \AA. 
\end{document}